\documentclass{elsart}

\usepackage{graphics}
\usepackage{graphicx}
\usepackage{multirow}

\usepackage{amssymb}

\journal{Physica D}

\begin{document}

\begin{frontmatter}



\title{Does the dynamics of sine-Gordon solitons predict active regions of DNA?}


\author[uc3m]{Sara Cuenda},
\ead{scuenda@math.uc3m.es}
\author[uc3m,bifi]{Angel S\'anchez},
\ead{anxo@math.uc3m.es}
\author[sevilla]{Niurka R. Quintero}
\ead{niurka@euler.us.es}

\address[uc3m]{Grupo Interdisciplinar de Sistemas Complejos (GISC) and
Departamento de Matem\'aticas, Universidad Carlos III de Madrid,
Avenida de la Universidad 30, 28911 Legan\'es, Madrid, Spain
}
\address[bifi]{Instituto de Biocomputaci\'on y F{\'\i}sica de Sistemas
Complejos (BIFI), Universidad de Zaragoza, 50009 Zaragoza, Spain
}
\address[sevilla]{Universidad de Sevilla,
Departamento de F{\'\i}sica Aplicada I,
E.U.P., Virgen de \'Africa 7, 41011 Sevilla, Spain
}

\begin{abstract}
% Text of abstract
In this work we analyze the possibility that soliton dynamics in a simple
nonlinear model allows functionally relevant predictions of the behaviour
of DNA. This suggestion was first put forward by Salerno [Phys.\ Rev.\ A
{\bf 44}, 5292 (1991)] by showing results indicating that sine-Gordon
kinks were set in motion at certain regions of a DNA sequence that include
promoters. We revisit that system and show that the observed behaviour
has nothing to do with promoters; on the contrary, it originates from the
bases at the boundary, which are not part of the studied genome. We
explain this phenomenology in terms of an effective potential for the
kink center. This is further extended to disprove recent claims that the
dynamics of kinks [Lenholm and H\"ornquist, Physica D {\bf 177}, 233 (2003)]
or breathers [Bashford, J.\ Biol.\ Phys.\ {\bf 32}, 27 (2006)] has
functional significance. We conclude that no such information can be
extracted from this simple nonlinear model or its associated effective
potential.
\end{abstract}

\end{frontmatter}

% main text 
\section{Introduction} 

Nonlinear models supporting coherent excitations appear in many fields of
science since the pioneering discoveries by Fermi, Pasta and Ulam \cite{fermi}
more than 50 years ago. The success of this approach in modeling complex systems
has encouraged its application in other fields. That is
the case of biology, where nonlinear models were widely applied in many
subjects, such as in the study of  the DNA molecule (see, for example, 
\cite{davydov,yakushevich,gaeta}).  To realize the relevance of these models it
should be noticed that, nowadays, the computational cost of molecular dynamics
for realistic models of DNA molecules with a few tens of base pairs allows 
simulation times up to tens of nanoseconds at most. Nonlinear models allow  the
study of such a complex system with very many  degrees of freedom by reducing
drastically this amount up to one degree of freedom per base pair, the most
relevant for the process under study. It goes without saying that the reduction
of a very complicated object such as the DNA duplex to a polymer formed by base
pairs, each one with just one degree of freedom (sometimes  a few more) helps
enormously the theoretical and computational study of this models. 
Nevertheless, although simplified, these models can yield important results. An
example of these models is the Peyrard-Bishop model of DNA \cite{peyrard}, which
achieved an important goal when describing the denaturation process of DNA in
terms of just the radial distance of the bases on each base pair 
\cite{nonlinear}.  

Among all these approaches we focus here on the work of Englander {\it et al.}
\cite{englander}, who introduced the sine-Gordon (sG) equation as a model for
DNA in
1980.  The existence of sG solitons in the DNA molecule has been surrounded by
controversy, as expected in a field were biology and physics  do not always meet
in a fruitful  way \cite{maddox,frank-ka}.  When Englander and co-workers
introduced the sG model of DNA,  they based their hypothesis on experimental
results that showed unexpectedly long lifetimes of open states of DNA duplexes 
\cite{hippel}.  In spite of the fact that, later, Gu\'eron {\it et al.}
 \cite{gueron}
found  more reasonable lifetimes, smaller by one or two orders of magnitude
than the ones reported in previous works, a vast amount of literature is still
based on Englander model. In this context, the  aim of this work is to  analyze
in depth part of the literature related to the work of Englander {\it et al.},
providing new results that give insight into a number of important questions.
Specifically, we will study the relation between  the dynamics of sG solitons
and the position of promoters in the genome of the bacteriophage T7. This line
of work began with Salerno \cite{salerno1,salerno2,kivshar} at the  beginning of
the 90's and was  subsequently continued in several works 
\cite{anxo,danish,sara1,sara2,bashford}. We stress that this is a very important
issue: Indeed, if the Englander model behaviour could be connected to
functionally relevant positions in the sequence, it would provide a cheap and
efficient tool for genomics. Although claims in this direction have been
recently presented \cite{bashford}, the main result of the present work is that,
unfortunately, such a connection cannot be substantiated.

The structure of the paper is as follows. In section \ref{sec:salerno}  we
discuss the methodology and the results of the two first papers about
this issue \cite{salerno1,salerno2} in terms of the effective potential 
introduced by Salerno in collaboration with Kivshar in \cite{kivshar}. In
section \ref{sec:t7} we describe the main features of the promoters of
the T7 genome, and analyze the simulation results of the work of Lennholm 
and H\"ornquist
\cite{danish} in terms of the effective potential. In section 
\ref{sec:breathers} we discuss recent work about breathers in the sG 
model \cite{bashford}. Finally, section 
\ref{sec:con} concludes the paper by summarizing our main results and
their implications.

\section{Early work on $T7$: $A_1$, $A_0$ and $A_3$ promoters}
\label{sec:salerno}
More than a quarter of a century ago,
Englander and coworkers \cite{englander} introduced solitonic 
excitations into the DNA world
as an initial step towards understanding the stability of open segments
of DNA molecules \cite{hippel}. They suggested the well-known 
sG model, that describes the dynamics of a line of pendula in a
vertical gravitational field with torsional spring coupling between 
units, as an effective description of DNA molecules. In this way, 
the double-helix
is approximated by two parallel rods on which pendula (base pairs)
are attached, and bonding to the opposite base is represented by
a ``gravitational'' potential of each pendulum. Calling $\phi_i$
the twist angle of the $i$-th base, this model has static soliton (kink) solutions 
given by
\begin{equation}
\label{eq:kink}
\phi_i=4\arctan\left(\mathrm{e}^{ai}\right),
\end{equation}
valid for $a\ll 1$, where the continuum approximation applies. In
equation (\ref{eq:kink}), $a$ is a dimensionless parameter representing
the parameters of the model, and acts as an effective discretization
parameter of the continuum sG problem. 
In spite of such a great oversimplification of
the real problem, the model contained the main feature of breaking 
a bond around $\phi=0$. In addition to this, the results were
consistent with available data \cite{hippel} although Englander {\it et al.}
were aware of the
lack of evidences of solitonic excitations.

Salerno, in his pioneering and interesting work
\cite{salerno1}, tried to find a relation between 
relevant sites in the $T7$ genome and the dynamics of sG kinks
moving along the inhomogenous DNA sequence under study.
The main difference with respect to previous works was
the introduction of the inhomogeneity of the sequence in the
model.
To do so, he took the static kink solution (\ref{eq:kink}),
with center at $n_0$,
and used it as initial condition of the equations of motion
of the discrete, {\em inhomogeneous} sG (or Englander) model, 
\begin{equation}
\label{eq:motion}
\ddot\phi_i=\phi_{i+1}-2\phi_i+\phi_{i-1}-q_i\sin\phi_i,
\end{equation}
$q_i$ being the parameter that carries all the information
of the sequence under study. It is defined as $q_i=\beta\lambda_i/K$,
where $K$ is the torsional spring constant between consecutive
bases, $\beta$ is the energy of a hydrogen bond and $\lambda_i$
is the number of hydrogen bonds in a base pair, which is $\lambda_i=2$
for $AT$ base pairs and $\lambda_i=3$ for $CG$ base pairs.
Considered as a discrete version of the continuum sG equation,
the effective discretization of the lattice used 
in \cite{salerno1} was $a=\bar q\,^{1/2}$, where
$\bar q=\frac{1}{N}\sum_{i=1}^Nq_i$ ($N$ being the number of bases
of the sequence). This value is
around $a\simeq 0.07$, which is small enough
to avoid spurious discretization effects when numerically integrating Eq. 
(\ref{eq:motion}). In fact, taking Eq. (\ref{eq:kink}) as an {\em Ansatz}
in Eq. (\ref{eq:motion}) was a good choice, as the kink is a very robust
object even in inhomogeneous sequences and its center can be well defined
by interpolating the position where $\phi=\pi$ \cite{sara1,sara2}.

Once the model was defined, Salerno built a sequence $\{q_i\}$ 
to introduce it in (\ref{eq:motion}). He was interested in the genomic
sequence of the $T7$ $A_1$ promoter but, instead of using the original
DNA sequence, he built a ``synthetic'' one from the original. We will review
all the details of this process as this will be the key to understand 
the results of \cite{salerno1}. He took a sequence $S$ of 168 bases
containing the so-called $A_1$ promoter (further details on $T7$ promoters
will be given in the next section)
which corresponds to base pairs (BP) from 378 to 545 of the actual 
$T7$ genomic sequence, and
built a longer sequence of 1000 bases, that we will call $S'$, 
to prevent the influence of boundary conditions on the 
dynamics of kinks:
\begin{equation}
\label{eq:seq}
S'=S(1,5)+8S(1,50)+S(1,168)+15S(141,168)+S(162,168).
\end{equation}
In this way, the 168 bases sequence $S$ would remain in the center of
the new sequence $S'$, with the transcription start site located in BP 526
and the promoter sequence going from BP 509 to BP 531,
far from the limits of the lattice. Therefore, reflective boundary 
conditions could be safely used in the numerical simulations.
We will return to this issue when discussing the results.

As was known that the RNA polymerase could bind to DNA in the region
of $S$ going from BP 51 to BP 140 (going from
BP 455 to BP 545 in $S'$), the expectation was that 
this region should be dynamically active.
Hence, in \cite{salerno1} several integrations of Eq. (\ref{eq:motion}) 
were carried out with
the initial position of the static kink in a variety of sites inside
the promoter region and the behaviour of the kinks was studied as a function
of their starting position. The results were the following:
For initial positions in $S'$ from BP 415 to BP 505 the kink remained static or
with small oscillations around the starting point. For BP 510, the
kink acquired a velocity $v=0.18$ towards the left, 
was reflected without loss of energy
at the left boundary and reflected again at the promoter region with
velocity $v=0.18$. This behaviour was enhanced when the initial position
was increased from BP 510 to BP 535, where the kink also reached 
the maximum velocity $v=0.3$. Beyond this point this dynamical behaviour 
was drastically reduced. For BP 540 the kink acquired a small velocity
($v\simeq 0.08$) towards the right, and for BP 555 the kink simply
remained at rest. The dynamics of a kink with
initial velocity $v=0.3$ towards the left was also studied, 
starting from BP 900; it was found that the
soliton was accelerated when it traveled from right to left through the
central region, then reflected at the left end of the sequence and
decelerated when traveled in the opposite direction. 
It was concluded in \cite{salerno1} 
that these results showed the existence of a dynamically 
``active'' region going from BP 510 to BP 540 inside the $T7$ $A_1$
promoter that could explain the functioning of DNA promoters as 
energetic activators of the RNA polymerase transport process.

\begin{figure}
\begin{center}
\includegraphics[width=60mm, angle=270]{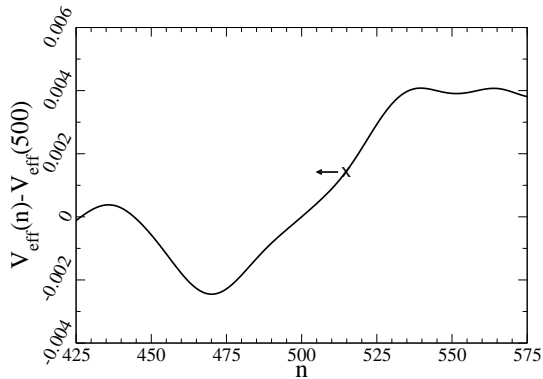}
\caption{\label{fig:salerno}
Effective potential for $a=0.07$ of sequence $S'$ 
from BP 425 to BP 605. The only difference with the one represented 
in \cite{kivshar} is that, in the latter, an additional average of the
potential over a distance equal to $\bar{q}\:^{-1/2}$ was made.}
\end{center}
\end{figure}

In a subsequent paper, Salerno and Kivshar \cite{kivshar}
introduced the effective potential in order to explain
the behaviour of these objects when moving in a inhomogenous sequence. The
idea is that kink robustness
allows to approximate their dynamics, even though they are extended
objects, as if they were point-like particules moving along a one-dimensional 
potential, given by
\begin{equation}\label{eq:poteff}
V_{\mathrm{eff}}(n)=\frac{\sum_{m}(\bar q+q_m){\mathrm{sech}}^2 (a(m-n))}
{2\sum_{m}{\mathrm{sech}}^2 (a(m-n))}.
\end{equation}
Recently this approach has been shown to give good results for Fibonacci 
\cite{anxo} and DNA sequences other than the $T7$ one
\cite{sara1,sara2}. By ``good results'' we mean that the dynamics of the 
kink in Eq. (\ref{eq:motion}) and that of the particle in the effective
potential (\ref{eq:poteff}) can by aligned, in the sense that trajectories
are semiquantitatively similar, equilibrium 
points for the kink correspond to minima
of the potential, and so on. 
This was also the case with the effective potential introduced
in \cite{kivshar}: This paper reported the agreement of the direction of motion
of the kinks according to the effective potential curve corresponding to
the sequence $S'$, plotted from BP 425 to BP 605 (see Fig. \ref{fig:salerno}).
As can be seen from the figure, there is indeed a good
correspondence between the effective potential and the simulation results
summarized above.

\begin{figure}
\begin{center}
\includegraphics[width=60mm, angle=270]{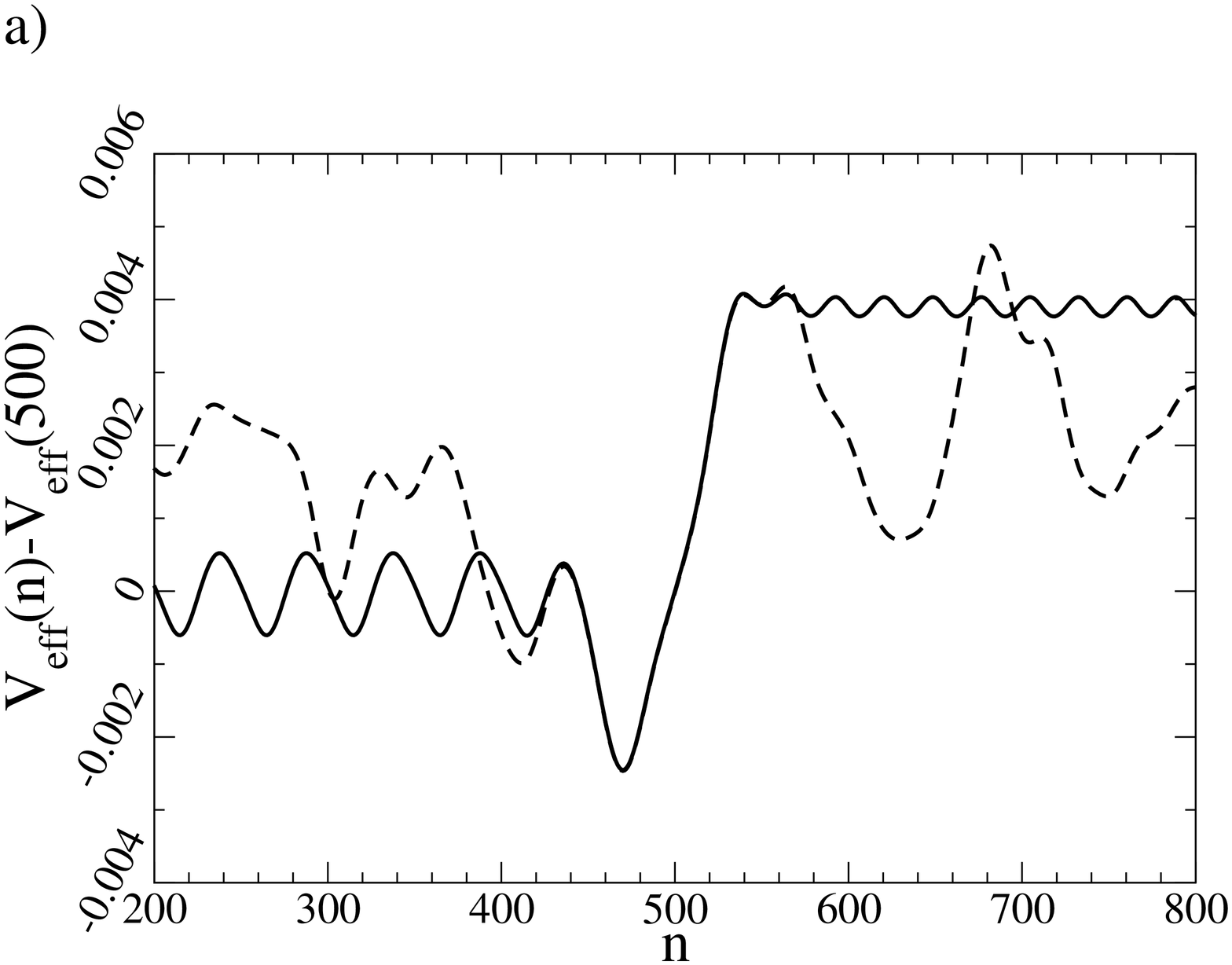}
\includegraphics[width=60mm, angle=270]{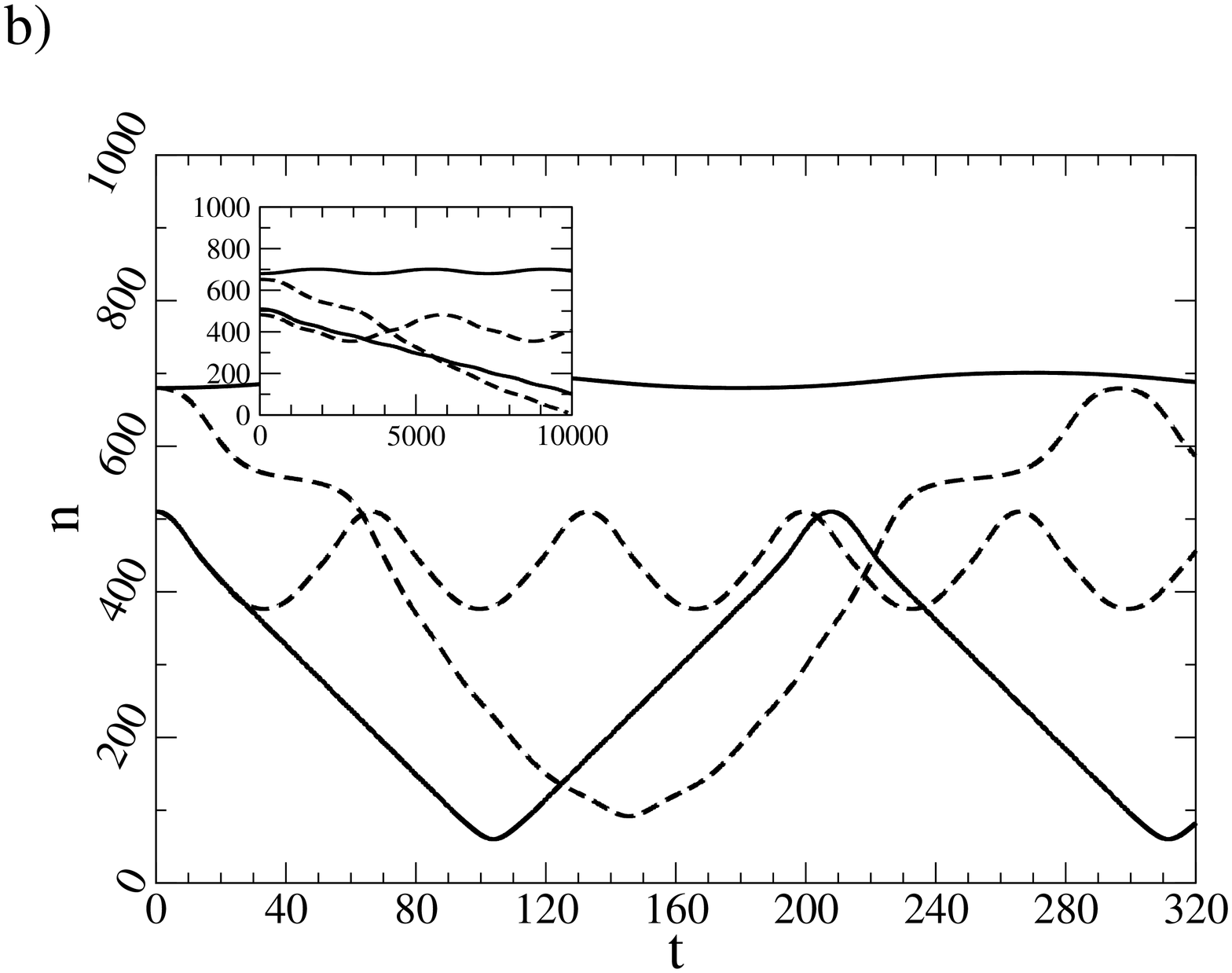}
\caption{\label{fig:a1}
(a)
Effective potential for $a=0.07$ 
around the $T7$ $A_1$ promoter for the synthetic
sequence $S'$ (solid line) and for the original genome sequence  
(dashed line). The potential of the original sequence has been 
shifted in the horizontal and vertical axis to make it coincide
with the one of the 168 bases sequence $S$ in $S'$.
(b) 
Dynamics of the center of two kinks (calculated by interpolating
the position where $\phi=\pi$), moving along $S'$ sequence (solid lines)
and the real $T7$ genome sequence (dashed lines), starting from
BP 510 and BP 680. Inset: point-like particle dynamics
starting from the same sites and moving according
to the corresponding effective potentials of (a). Although the
dynamics is scaled in time with respect to the kink dynamics,
the trajectories of each pair particle/kink are the same.
}
\end{center}
\end{figure}

\begin{figure}
\begin{center}
\includegraphics[width=60mm, angle=270]{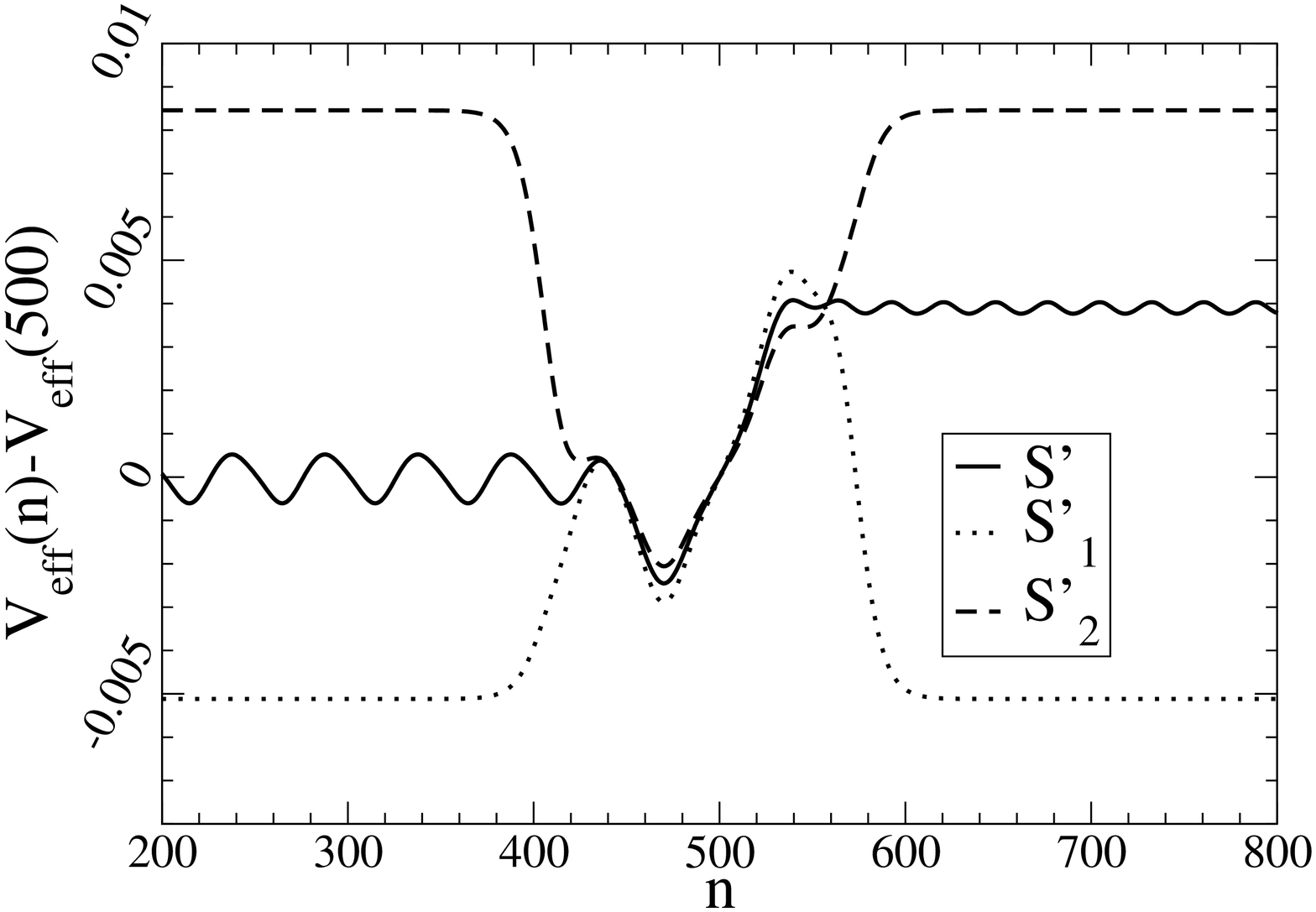}
\caption{\label{fig:a1_b}
Relevance of the end parts of the sequence:
Effective potential for the synthetic sequences $S'$,
$S'_1$ and $S'_2$ (see text for definitions).
}
\end{center}
\end{figure}

However, a more detailed analysis shows that this correspondence is not
enough to establish a relation between DNA promoters and dynamically 
``active'' regions. 
In Fig. \ref{fig:a1}a it is plotted the effective potential 
$V_{\mathrm{eff}}(n)$ (taking $V_{\mathrm{eff}}(500)$ as the origin
of energies)
for sequence $S'$ and for the original $T7$ genome sequence.
From this figure we immediately observe, on the one hand, 
that the positions of the peaks and
the wells of the effective potential of sequence $S'$ explain
very well the above kink dynamics results reported in \cite{salerno1}
in terms of a point-like particle; and, on the other hand, that 
the effective potential of the original $T7$ genome {\em far from the $A_1$
promoter} is very different from the one of the $S'$ sequence,
and hence the dynamics of kinks must be different, too. Fig.
\ref{fig:a1}b shows the dynamics of two kinks 
on the two sequences, $S'$ and the original $T7$ sequence.
It is clear that the dynamics behaviour of both sequences is largely different:
for instance, in the true potential
BP 510 should not be regarded as an active site whereas 
BP 680 should be regarded so. 
The comparison of the trajectories with those obtained from the effective
potential confirms the validity of this potential to describe the dynamics
of kinks (allowing for a difference in time scales, as in the point-like
particle description time units are arbitrary). This means that the 
effective potential is a correct description for both the real sequence
and $S'$, and therefore the differences between both of them are not an
artifact of this approximation.
These differences between the two potentials come from the periodic
sequences introduced in (\ref{eq:seq}), $8S(1,50)$ and $15S(141, 168)$, 
adjacent to the 168 nucleotide $S$ sequence. The AT/CG content in
the periodic sequences has an average value around which the effective
potential of these sites oscillates. 
As further evidence of the influence of the ends of the sequence,
in Fig. \ref{fig:a1_b} we show
the effective potential of sequences $S'$, $S'_1$ and $S'_2$, with
$S'_1=405\,A+S(1,168)+427\,A$ and $S'_2=405\,C+S(1,168)+427\,C$,
where $N\,A$ ($N\,C$) means $N$ consecutive sites with nucleotide 
$A$ ($C$). The effective potentials for kinks moving along $S'_1$ and $S'_2$
will lead to a very different dynamics from that described in \cite{salerno1}
and reported here,
although they all have the same central sequence of 168 nucleotides.

\begin{table}
\begin{center}
\begin{tabular}{|c|c|l|}
\hline
{\bf Promoter} & {\bf BP region} & {\bf Response}\\
\hline
\multirow{2}{*}{$A_1$} & 
   From 510 to 535 & Leftward propagation, strongest at 535.\\
   & 540 & Small velocity towards the right.\\
\hline
\multirow{2}{*}{$A_0$ (or $D$)} & 
   From 530 to 540 & Leftward propagation, strongest at 540.\\
   & From 543 to 555 & Rightward propagation, strongest at 543.\\
\hline
$A_3$ & From 435 to 460 & Rightward propagation.\\
\hline
\end{tabular}
\vspace{3mm}
\caption{\label{tab:dyn}
Summary of the dynamical results for kinks moving in the 
synthetic sequences $S'$ obtained from $A_1$, 
$A_0$ (also called $D$) and $A_3$ promoters in \cite{salerno2}.}
\end{center}
\end{table}

\begin{figure}
\begin{center}
\includegraphics[width=60mm, angle=270]{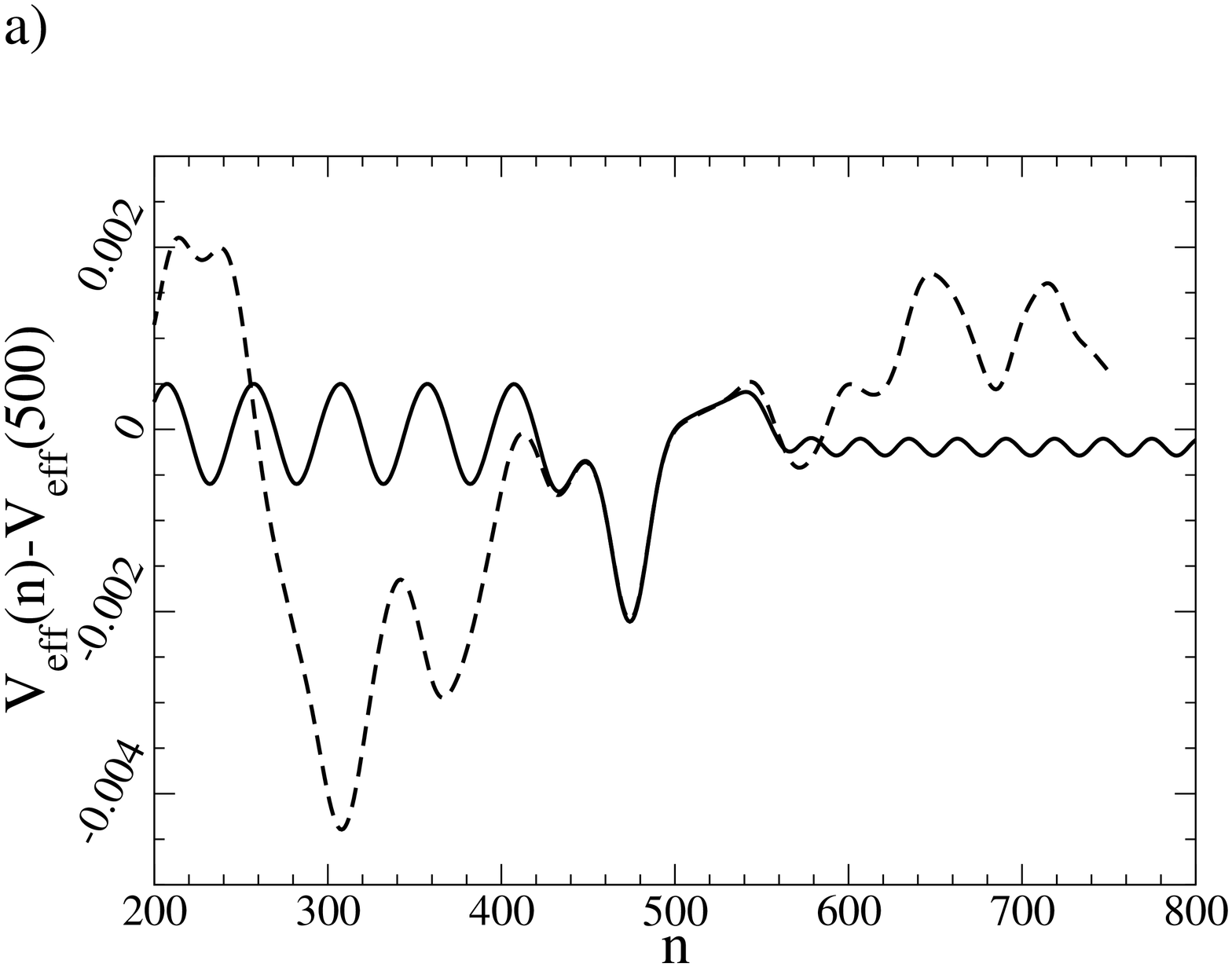}
\includegraphics[width=60mm, angle=270]{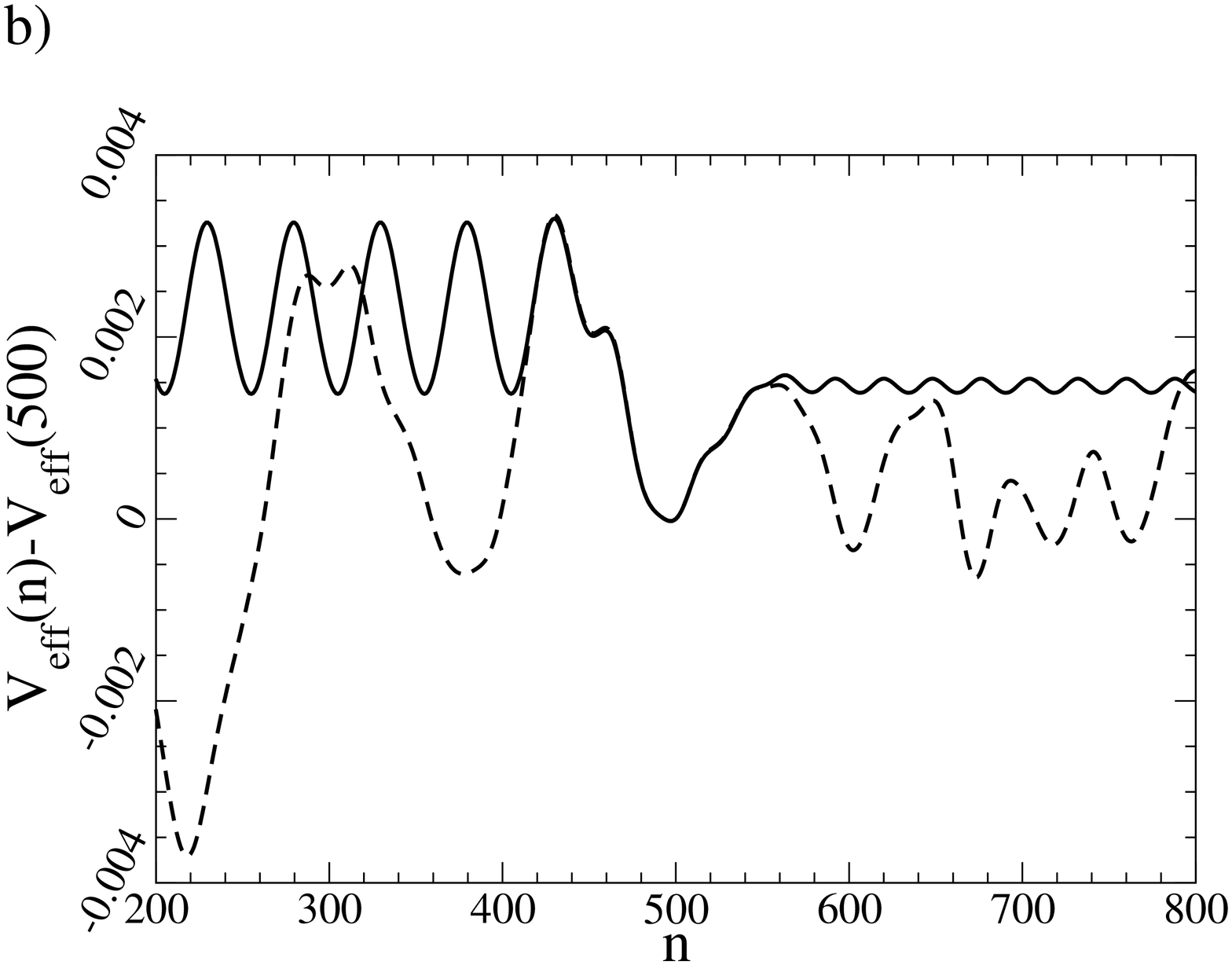}
\caption{\label{fig:a03}
(a)
Efective potentials for $a=0.07$ of the synthetic sequence $S'$ (solid line) 
and for the original genome sequence (dashed line) around $T7$ $A_0$
(also called $D$) promoter. The sequence used in \cite{salerno2} corresponds
to the transcription strand (this promoter activates transcription
leftwards, and therefore it uses the complementary strand of the
sequence usually showed \cite{dunn,ncbi}) in the transcription order.
Therefore, the original sequence has been written in reverse order in 
order to obtain its potential, and then shifted in the horizontal and
vertical axis to make it coincide with the potential of the $S$ sequence in
the synthetic $S'$ sequence.
(b)
Efective potentials for $a=0.07$ of the synthetic sequence $S'$ (solid line) 
and for the original genome sequence (dashed line) around $T7$ $A_3$ promoter.
The potential of the original $T7$ sequence has been shifted in order
to make it coincide with the potential of the $S$ sequence in
the synthetic $S'$ sequence.
}
\end{center}
\end{figure}

In \cite{salerno2} the same methodology developed in \cite{salerno1} was used
to analyze another two $T7$ 
promoters, namely $A_0$ (also called $D$) and $A_3$, and
similar results were obtained (see Table \ref{tab:dyn}). Figs. \ref{fig:a03}a
and \ref{fig:a03}b show the effective potential of the synthetic
sequences built by
Salerno from the genomic 168 nucleotide sequences and the effective potentials
of the real T7 sequences around the promoters. Again, the effective 
potentials of the synthetic sequences describe the results of the
dynamics summarized in Table \ref{tab:dyn}, but differ from
the effective potentials of the real genomic sequences,
yielding different dynamics. For instance,
according to Fig. \ref{fig:a03}a, a kink starting from BP 245 in the real
genomic sequence around $A_0$ promoter
would reach the right end of the sequence, instead of
oscillating around the initial starting position, as it would do in
the corresponding $S'$ sequence.

We note that in \cite{salerno2} it was argued that, as the initial static
soliton was always well inside the original 168 base sequences, then the
flanking regions used to prolog the chain played no role in the dynamical 
effects described. However, we have just shown how important they are
when the kink moves towards them.
Therefore, we are forced to conclude that the results in
\cite{salerno1,salerno2} 
are highly dependent on the construction of sequences $S'$, and that 
when the original
$T7$ genome sequence is used instead then the promoter regions cannot be 
considered ``active'' or ``special'' regions anymore. As we have seen,
other regions close to, but different from, the promoters may be even more
``active'' in the sense of inducing kink motion; conversely, some active
regions in the synthetic sequence lose this character in the real genome.

\section{Subsequent developments: full $T7$ genome}
\label{sec:t7}

Following the interesting proposal of Salerno, namely
the putative relation between $T7$ promoters and the dynamics of solitons
moving along inhomogeneous genomic sequences, further research
intended to shed further light on this question \cite{danish}. 
The main contribution of this sequel is that
the sequences used were real genomic sequences and, in addition, that the
whole $T7$ genome was studied.

In the research reported in \cite{danish}, 
Lennholm and H\"ornquist measured the maximum distance (in either 
direction) reached by initially static kinks starting from
each of the sites of the whole sequence of the $T7$ genome.
They also took the 24 promoters of the $T7$ genome (except the first 
and the last ones to avoid boundary effects), studied the results 
obtained for positions going from -4 to -1 of each promoter and compared these
results with the results of the whole genome. The aim of this analysis
was to find whether the RNA polymerase melting region (the one going
from -4 to -1 in each promoter) acts as a dynamically ``active'' region
as proposed by Salerno, or behaves in the same way as the rest
of the nucleotides of the genome. In this respect, they did not found
relevant differences (see Fig. 1 of \cite{danish}). 
However, for every promoter they investigated the activity of the
first $n$ base pairs which are transcribed by RNA polymerase
and found that, for $n=20$,
the studied regions are more active than average with a significance
of more than five standard deviations (see Fig. 2 of \cite{danish}).
They did not give any biological interpretation of that results but,
in their conclusions, they suggested that a more quantitative relation
between kink motion and the effective potential should be established.

As already mentioned, this paper as well as 
our previous work \cite{anxo,sara1,sara2} has
proven the agreement between kink dynamics and effective potential.
Therefore, we can now study the whole $T7$ genome in terms
of this tool. To this end, we will review some of the properties
of the $T7$ promoters in order to set a methodology in the study of 
the effective potential in these regions.
The $T7$ phage genome is one of the most studied genomes 
since the whole genome sequence was found in 1983 \cite{dunn},
and few changes in the sequence have been reported since then \cite{ncbi}.
The reproductive cycle of the $T7$ phage is closely linked to the promoter
and gene distribution in the genome. When the $T7$ RNA is injected
inside a bacteria, like {\em E. coli}, the bacterial RNA polymerase starts
to produce mRNAs induced by three major promoters
from the  early region (or class I region) $A_1$, $A_2$
and $A_3$. A fourth major {\em E. coli} promoter, $A_0$ (also called $D$),
that would direct transcription leftward, and several minor {\em E. coli}
promoters function in vitro but have no known in vivo function.
Once the $T7$ phage has its own transcription machinery, late
mRNAs are produced by 15 promoters for $T7$ RNA polymerase distributed
across the right-most 85\% of the DNA (divided in class II and class III
region). There are also two $T7$ promoters
associated with possible origins of replication at the left and
right ends of $T7$ DNA.
The 23 base-pair consensus sequence for $T7$ promoters stretches 
from -17 to +6, where +1 is the transcription start site. This 
means that the nucleotides of the promoters are highly correlated
in these sites (although sometimes they are not strictly the same), 
but not in the 
rest of the sequence (we will come back to consensus sequences below).

\begin{figure}
\begin{center}
\includegraphics[width=53mm, angle=270]{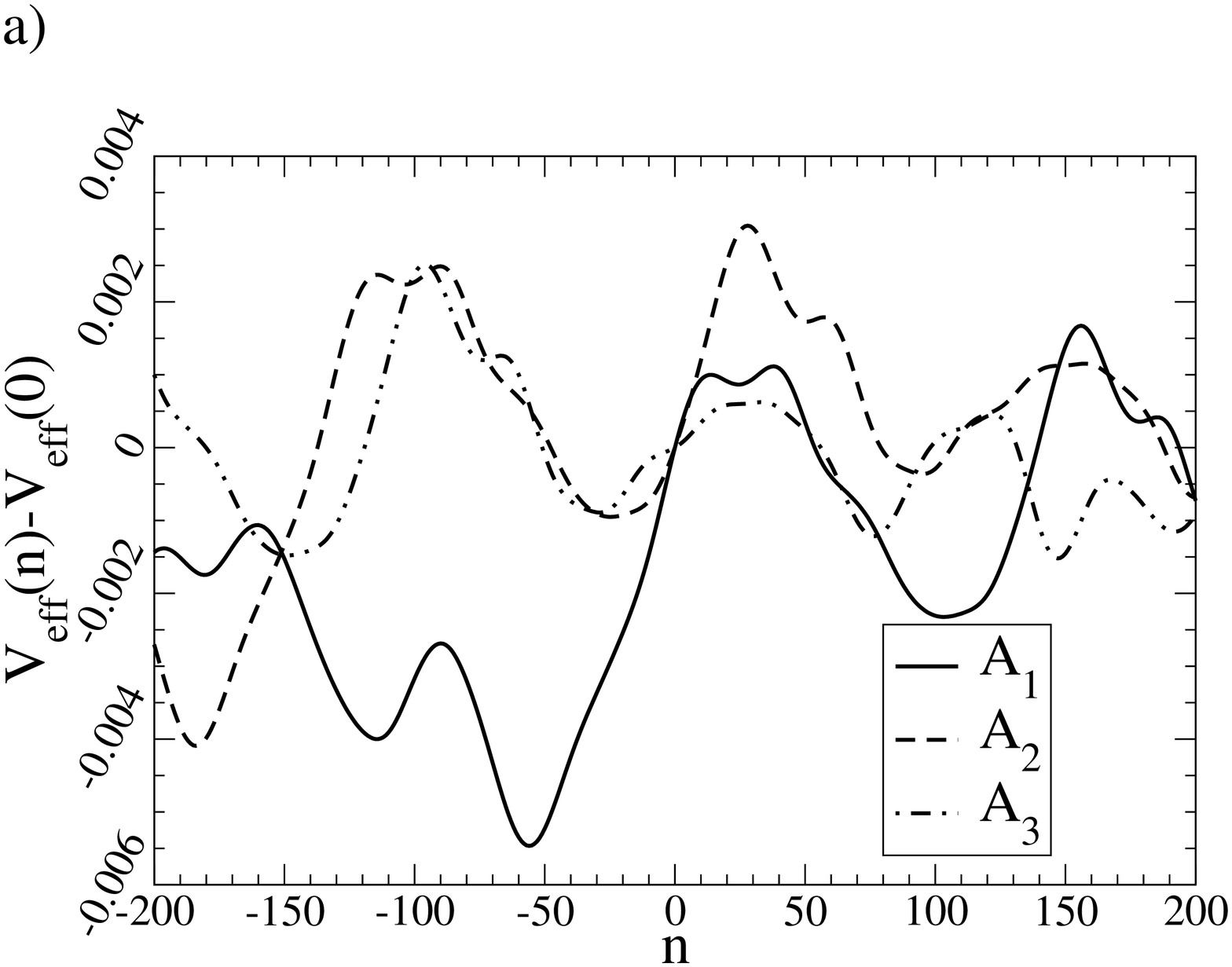}
\includegraphics[width=53mm, angle=270]{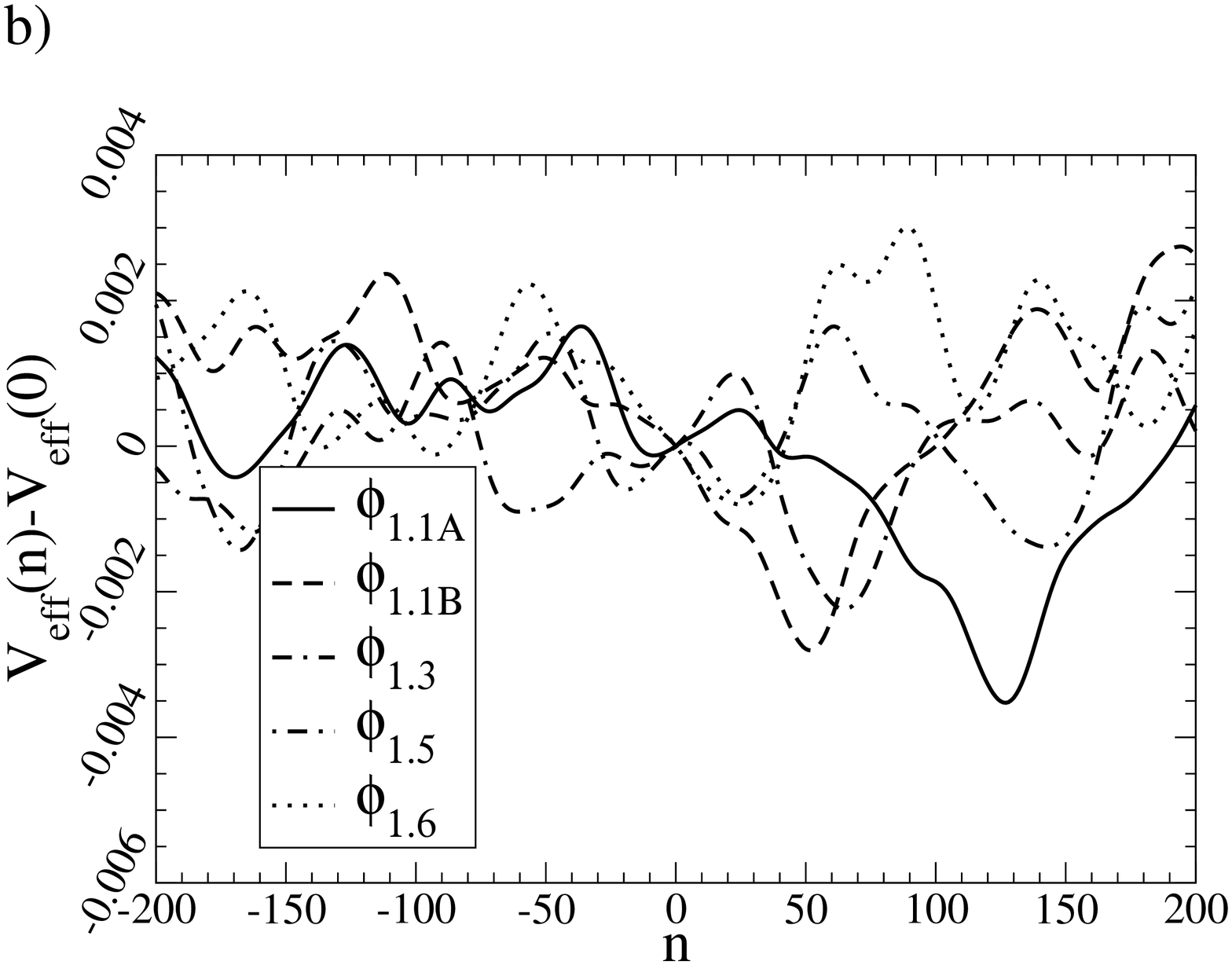}\\
\includegraphics[width=53mm, angle=270]{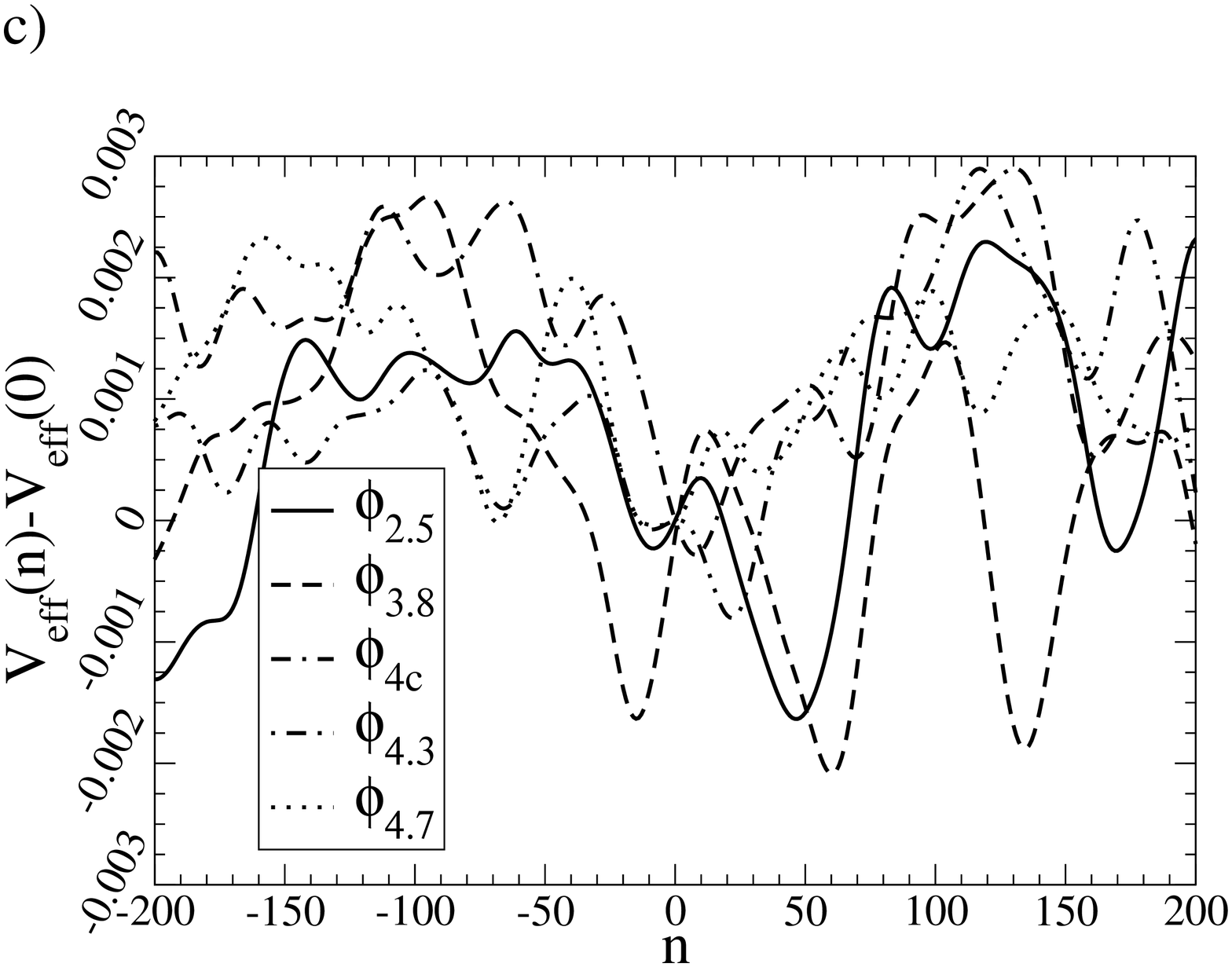}
\includegraphics[width=53mm, angle=270]{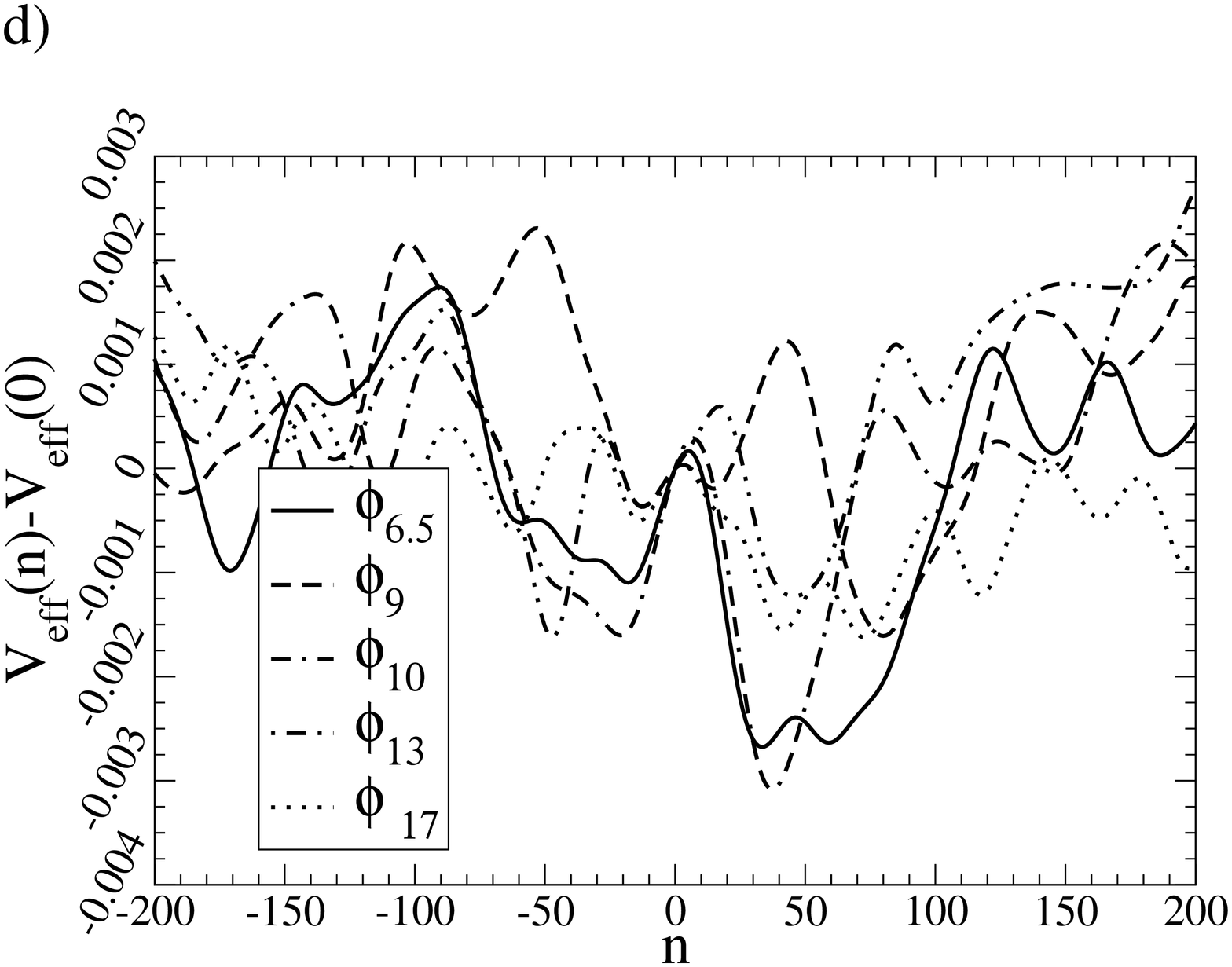}
\caption{\label{fig:promoters}
Effective potentials for $a=0.07$ around the transcription
start site of: (a) Early $A_1$, $A_2$ and $A_3$ E. coli promoters; 
(b) Late $\phi_{1.1A}$, $\phi_{1.1B}$, $\phi_{1.3}$, $\phi_{1.5}$ and 
$\phi_{1.6}$ $T7$ promoters; 
(c) Late $\phi_{2.5}$, $\phi_{3.8}$, $\phi_{4c}$, $\phi_{4.3}$ and 
$\phi_{4.7}$ $T7$ promoters; 
(d) Late $\phi_{6.5}$, $\phi_{9}$, $\phi_{10}$, $\phi_{13}$ and 
$\phi_{17}$ $T7$ promoters. Origins are referred to the transcription
start site in all cases.
}
\end{center}
\end{figure}

We can now go back to our main aim: We want to find out whether or not there is
some kind of pattern in the effective potential,
or a set of properties to be applied to all the promoters in the $T7$
genome that allow their identification among the rest of the genome.
To this end, we must keep in mind that the effective potential on each
site [see Eq. (\ref{eq:poteff})] is just a weighted average of the sequence 
around the site, with weight function $\mathrm{sech}^2(an)$. 
We can obtain an estimation of the resolution of the effective potential
reading frame by noticing that an error of about 10\% in
the effective potential is introduced
when truncating the sum in the weighted average (\ref{eq:poteff}) in
$\pm\Delta n=1.5a^{-1}$ around each $n$. If we consider that, for further 
sites, the contribution of the $q_m$ to the weighted average is negligible,
then the number of sites averaged when computing the effective potential
on each site goes as $\Delta n\simeq 3a^{-1}$.
This means that, for
$a\simeq 0.07$ (which is the approximate value of the discretization
as explained in section \ref{sec:salerno}), $\Delta n$ is about
40, a much lower resolution than the one needed to recognize the
23 base-pair consensus sequence in the effective potential. Therefore,
we conclude that the kink is too wide to allow us to check that
the same curve describes the effective potential of different promoters,
as was suggested in \cite{salerno1,salerno2,danish}.
On the other hand, we can increase the resolution of the effective potential 
by increasing the discretization $a$ until reaching $\Delta n=1$.
We could find then the consensus sequence repeated in the effective
potential around each promoter, but that would not give more information
than the consensus sequence itself, and the effective potential would
not be useful from a genomic point of view.

\begin{figure}
\begin{center}
\includegraphics[width=60mm, angle=270]{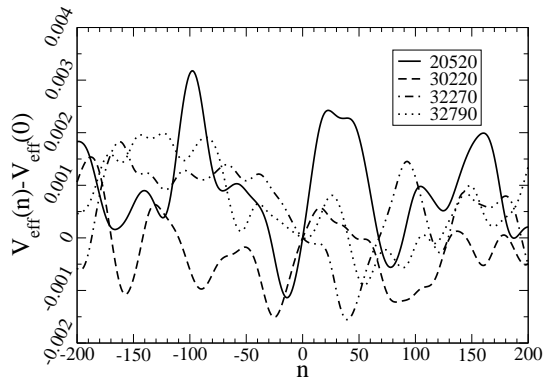}
\caption{\label{fig:other}
Effective potential for $a=0.07$ in different sites 
of the sequence, which may
look like the effective potential of promoters in Fig. \ref{fig:promoters} 
but which are not.
The origin of each sequence corresponds to the number in the legend box.
}
\end{center}
\end{figure}

To go beyond the previous theoretical discussion, we have computed the
effective potential for most of the $T7$ promoters for $a=0.07$.
In Fig. \ref{fig:promoters} we show the
effective potential of the three major {\em E. coli} RNA polymerase 
promoters (early promoters) and the
T7 RNA polymerase promoters (late promoters)
of $T7$ for $a=0.07$. Clearly there is no
``consensus effective potential'' that appears in all (neither in
most) of the promoters. If we were looking for more subtle properties that
might enclose all the 18 promoters or each subset of early and late promoters,
then we would be led to consider as promoters other regions
of the $T7$ genome which are not. As an example,
the effective potentials around
some other regions that are not promoters are plotted in Fig. \ref{fig:other},
and it is shown how alike they are to the ones of Fig. \ref{fig:promoters}.
Hence, we believe that the effective potential of kinks, and therefore the
dynamics of kinks in the inhomogenous sG model cannot explain the
initiation in the promoters of the transcription process in the $T7$ 
phage and, probably, in any other organism.

We now turn to the work in \cite{danish} in order to understand
the difference with the conclusions reported there.
The research in \cite{danish} is certainly interesting because,
as we already said, it is the first time
that the whole genomic sequence of the $T7$ phage is taken into account.
However, we believe that their methodology is not appropriate for the 
case under study,
as the statistical analysis of kink dynamics does not give conclusive
results. For instance, a graph with the furthest position reached in 
the sequence in terms of the initial position from which the kink started
to move would have yielded different results from the ones reported
in \cite{salerno1,salerno2} and the work would have been more 
conclusive. In addition, we note that the direction of motion
of the kinks was not recorded and therefore it cannot be assessed whether
or not the ``activity'' of those regions agrees with the transcription
direction.
We therefore conclude that an individual study of each promoter is needed
if functionally relevant places are to be found.
This individual study is what we have presented here and we believe that
the conclusion is clear: The effective potential shows no signature of
the promoters. Having verified this, in the next section we show that, 
if a detailed study
of promoters is done, it must be over all the promoters of the phage
in order to be conclusive. 

\section{sG breathers}
\label{sec:breathers}

In this last section, we consider yet another recent approach to sG soliton 
dynamics that followed the steps of \cite{salerno1,salerno2,kivshar}
but using breathers instead of kinks \cite{bashford}. 
In this case, the author studied
RNA polymerase recognition of specific binding sites by
comparing breathers
to localised deformations of the DNA duplex when the RNA polymerase
slides on its major groove. To this end he constructed a potential for sG
breathers following the steps of \cite{kivshar}, but using as
{\em Ansatz} a discretized breather, given by
\begin{equation}\label{eq:breather}
\phi_{\mathrm{br},n}(t)=4\tan^{-1}\left(\tan\mu\frac{
\sin(t\ \bar q\,^{1/2}\cos\mu)}{\cosh(n\ \bar q\,^{1/2}\sin\mu)}\right),
\end{equation}
where $\mu$ is related to the intrinsic frecuency of the breather.
The potential obtained in \cite{bashford} is the following:
\begin{equation}
\label{eq:potbreath}
V_{\mathrm{br}}(n)=4\tan^2\mu\sum_m
\frac{(\bar q+q_m)\cosh(a(m-n))}{\left(\tan^2\mu+\cosh^2(a(m-n))\right)^{3/2}},
\end{equation}
with $a=\bar q\,^{1/2}\sin\mu\simeq 0.04$. The main difference between this
potential and the one obtained in \cite{kivshar} is that breathers
defined by
(\ref{eq:breather}) are not static solutions of the sG model. This means,
on the one hand, that the kinetic term of the sG Hamiltonian (that we
do not write here) has two extra terms when deriving 
$\phi_{\mathrm{br},n-n_0(t)}$ with respect to time and squaring it, and that the
potential term obtained depends explicitly on time. These problems may
be solved by moving one of the extra terms from the kinetic to the
potential term, and then integrating in time over a period.
Another important difference with \cite{kivshar} is that 
kinks are very robust objects
that behave very well in the discrete sG model, even for inhomogeneous
sequences, and that is why they can be expressed in terms of its center. 
Breathers, however, are very unstable in the homogeneous, discrete sG model,
and it is to be expected that they are
even more so on inhomogeneous sequences. Therefore, we do not think that
the potential for breathers may describe accurately and/or for long times
the dynamics of an initially static breather. 
The construction of the sG potential for breathers is, therefore, not
as straightforward as the one for kinks in \cite{kivshar},
and this must be taken into account when analyzing the results.

\begin{figure}
\begin{center}
\includegraphics[width=60mm, angle=270]{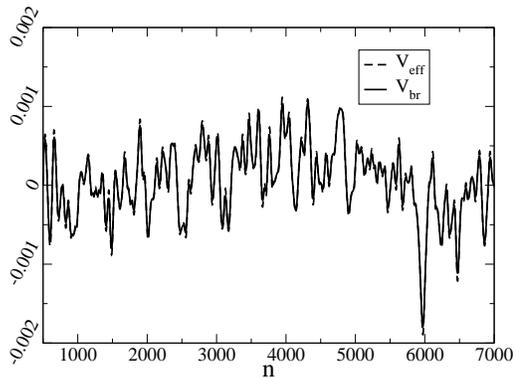}
\caption{\label{fig:compare}
Potential $V_{\mathrm{br}}$ (\ref{eq:potbreath}) for breathers (solid line) and 
$V_{\mathrm{eff}}$ (\ref{eq:poteff}) for kinks 
(dashed line) for the $T7$ genome region going from BP 500 to BP 7000
for $a=0.04$ and $\mu=\pi/6$ (which corresponds to the values
used in the figures of \cite{bashford}). The scale of $V_{\mathrm{br}}$ 
has been divided by $50$ in order to make it fit with the scale of
$V_{\mathrm{eff}}$. No vertical shifting was made in any potential.
This region of the genome contains seven promoters.
}
\end{center}
\end{figure}

After the breather potential (\ref{eq:potbreath}) was constructed in
\cite{bashford}, it was used to analyze the early region
of the $T7$ genome and a particular region of the $T5$ phage.
It was suggested, among other things, that there is a
correlation between deep wells of (\ref{eq:potbreath})
and promoters in the early region
and class III $T7$ promoters. Another relation between deep wells and
transcription terminators was also suggested.
We can now apply the results obtained in section
\ref{sec:t7} to the claims in \cite{bashford} by noticing 
how alike the weight functions of (\ref{eq:poteff}) and (\ref{eq:potbreath})
they are for $\tan\mu<1$ (which is the case, according to \cite{bashford}). 
Therefore,
the structure of peaks and wells is very similar in both cases, as
shown in Fig. \ref{fig:compare}, and we can thus extend the conclusions
of section \ref{sec:t7}: Even if the potential for sG breathers 
works in a similar way as the effective potential for sG kinks, it is not
enough to explain the transcription process of RNA polymerase.
For instance, as shown in \cite{bashford}, there are deep wells in the potential
for sG breathers near some of 
the promoters of the $T7$ genome. However, there are other promoters
(class II) which are not near any deep well, and also deep wells which are
not near promoters (like the ones found in \cite{bashford} near terminators).
When loosing the constraints in order to take into account not so deep wells
which are near promoters, then many other wells far from promoters should be
considered, too.
We therefore conclude that there is no special characteristic in the
potential for breathers that allow the identification of promoters from
the rest of the genome, by simply looking at the effective potential.

\section{Conclusions}
\label{sec:con}
The Englander model was introduced in \cite{englander} to explain
long life times on open states of DNA duplexes \cite{hippel} by
means of the well known nonlinear sG model. 
Subsequently, research on the sG model 
led to suggestions of a relation between functionally
relevant positions in the sequence with dynamical properties of sG
solitons. In section \ref{sec:salerno} we showed that the results
of kink dynamics moving along inhomogenous sequences developed 
in \cite{salerno1,salerno2} depend highly on the sequence under study.
In order to achieve this we used the effective potential for sG kinks moving on
inhomogeneous sequences, introduced in \cite{kivshar}.  
Applied to the sequences used in \cite{salerno1,salerno2} and
to the corresponding real genomic sequences of the $T7$ phage,
we observed important differences between both potentials.
Differences came from the end parts of the analyzed sequences, which
were a priori assumed not to have any role.
With these findings and taking into account the good results already
obtained for the particle-like approximation of sG kinks moving
along inhomogeneous sequences \cite{anxo,sara1,sara2}, 
we concluded
that early promoter regions of the $T7$ genome cannot be considered
dinamically ``active''. In section \ref{sec:t7}, addressing the question 
posed in
\cite{danish},  we searched for patterns that could differentiate 
the dynamics of kinks starting from $T7$ promoters from kinks starting 
from the rest of the genomic sequence. Again, we used the effective potential,
this time applied to the whole genomic sequence of the $T7$ phage.
Comparing the curves obtained for the 18 major promoters of the phage
among them and also with other non-promoter regions led us to think
that there was no special properties of the effective potential around
promoter regions, and therefore that the dynamics of kinks moving from
these regions was the same as in other genomic regions. Finally, in
section \ref{sec:breathers} we applied the same arguments and also 
reviewed the problems of the potential for breathers in order to
demonstrate that the potential for sG breathers obtained in \cite{bashford}
can not be used to differentiate promoter regions in the
genomic sequence. From all this evidence, we can confidently claim that
neither the sG model nor its description in terms of the effective
potential give hints about functionally relevant sites of DNA sequences.
We stress that this claim is about the sG model and the dynamics of
its solitons. Statistical mechanics approaches are also being studied
with some success \cite{usheva,kalosakas,rapti,cuesta} but that is a completely
different approach.

It is important to extend this discussion to include its biological
implications.
The relation of deep wells and functioning sites of DNA can now be discussed
in terms of properties of bacterial promoters
\cite{elibro,busby}. Bacterial RNA polymerase is a multisubunit  complex.  A
detachable subunit, called $\sigma$ factor, is responsible for reading the
promoters, which are the 
signals enconded in the DNA that tell it where to begin transcribing. 
Most bacteria contain multiple $\sigma$ factors that enable the
recognition of different sets of promoters.  A comparison of many different
bacterial promoters reveals that they are heterogeneous in the DNA sequence.
However, they all  contain related sequences that reflect on mechanical and 
electrostatic properties of the DNA double helix that are recognized by the
$\sigma$ factor. These common features are often summarized in the form of a
consensus sequence, which serves as a summary or ``average'' of a large number
of individual nucleotide sequences. The precise sequence determines the strength
(or number of initiation events per unit time) of each promoter. However, 
although
the $\sigma$ factor is needed in the transcription initiation, other elements
can bind to RNA polymerase to regulate the transcription of specific promoters,
like the $\alpha$ subunits. Another important group of proteins that recognizes
and binds to promoters are transcription factors. These proteins act as
regulatory elements that control transcription initiation and bind to specific
sequences. This summary of regulatory elements of transcription initiation in
procaryotes reveals the intrinsic  complexity of the sequences of promoters. In
the case of eukaryotic regulation the complexity increases too much to try to
summarize it in this paragraph, and we will just refer to the counter-intuitive
fact that specific CG-rich
promoters that are found in yeast \cite{marin}. Therefore, we conclude that,
although deep wells in the potential for  sG kinks or
breathers are correlated with AT-rich
regions, they are not enough to recognize such complex structures as
promoters, and it is only natural that the dynamics of these simple
excitations cannot capture the mechanisms of promoter function.

\begin{ack}
We thank Antonio Mar{\'\i}n for patient and helpful discussions about promoters.
This work has been supported by the
Ministerio de Educaci\'on y Ciencia of Spain through grants
BFM2003-07749-C05-01, FIS2004-01001, NAN2004-09087-C03-03,
FIS2005-973, by Comunidad de Madrid grant SIMUMAT-CM and
by the Junta de Andaluc\'ia under projects 00481 and FQM-0207. S.C.
is supported by a fellowship from the Consejer{\'\i}a de
Edu\-ca\-ci\'on de la Comunidad de Madrid and the Fondo
Social Europeo. 

This work originated from discussions at a Summer School
in 2005 at Baeza, Spain. We are grateful to Renato \'Alvarez Nodarse
for creating that nice atmosphere.
\end{ack}

% \appendix

% \section{}
% \label{}

\end{document}